\documentclass[11pt]{article}
\usepackage{moriond,epsfig}

\def\NIMA{{\em Nucl. Instrum. Methods} A}

\def\PRL{{\em Phys. Rev. Lett.} }
\def\PRD{{\em Phys. Rev.} D}

\def\ra{\rightarrow}

\def\be{\begin{equation}}
\def\ee{\end{equation}}
\def\bea{\begin{eqnarray}}
\def\eea{\end{eqnarray}}

\begin{document}
\vspace*{4cm}
\title{$\mathbf{B^{0}_{s}}$ MIXING, 
LIFETIME DIFFERENCE AND RARE DECAYS AT THE TEVATRON}

\author{ S. BURDIN (for the CDF and D\O\ collaborations)}

\address{Fermi National Accelerator Laboratory,
Batavia, Illinois 60510, USA}

\maketitle\abstracts{
Recent results on $B^{0}_{s}$ mixing, lifetime difference and rare decays 
obtained by the CDF and D\O\ collaborations using the data samples 
collected at 
the Tevatron Collider in the period 2002~--~2005 are presented. 
}

\section{Introduction}

 Run II at the Tevatron Collider started in 2001. The CDF and D\O\ 
experiments successfully collect data since that time. Until March 2005
each experiment recorded data corresponding to an integrated
luminosity of about 600~pb$^{-1}$. The analyses described in this paper
are based on samples corresponding to luminosity from 170 to 
510~pb$^{-1}$. 

 An important and currently unique capability of the Tevatron is production
of all species of $B$ hadrons. 
In particular, $B_{s}$ meson studies 
together with the precise $B_{d}$ measurements 
at $B$ factories~\cite{bfac} allow the 
overconstraint of the CKM matrix elements or the measurement 
of their ratios that are free from 
some theoretical uncertainties.

\section{$\mathbf{B^{0}_{s}}$ mixing and lifetime difference}

 The mass eigenstates of the two neutral $B-\overline{B}$ systems, 
$B^{0}_{d}-\overline{B^{0}_{d}}$ and $B^{0}_{s}-\overline{B^{0}_{s}}$, do not coincide with the 
corresponding flavor states but can be expressed as their linear combinations. 
 The $B_{d(s)}$ mass eigenstates have different masses and probably lifetimes 
leading to a property of $B^{0}_{d}$ and $B^{0}_{s}$ mesons to change flavor 
and transform into their antiparticles. This phenomenon is called oscillation or mixing. 
The oscillation frequency is proportional to the mass difference $\Delta m_{d(s)}$.
 The $B^{0}_{d}$ oscillation frequency is very well measured with the highest  
accuracy achieved at the BABAR and BELLE experiments~\cite{bd_freq}. A comparison of 
this frequency with the CKM matrix element $V_{td}$ involves hadronic parameters that
can be calculated from Lattice QCD and currently have large uncertainties ($\sim 15\%$)~\cite{latQCD}.
Measurement of the 
$B^{0}_{s}$ oscillation frequency, which is at least 30 times higher than that of the $B^{0}_{d}$ meson,
will allow to determine the $\Delta m_{d}/ \Delta m_{s}$ ratio which is related to the 
CKM matrix elements $V_{td}/V_{ts}$ through the hadronic parameters ratio $\xi$. The $\xi$ parameter
is known with much less uncertainty ($\sim 5\%$)~\cite{latQCD}.

\begin{figure}
\begin{center}
\epsfig{figure=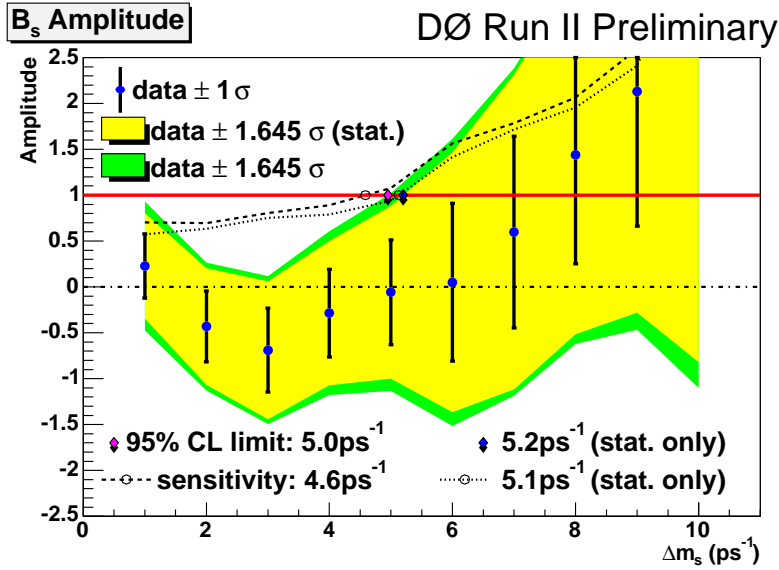,height=3in}
\caption{D\O\ Collaboration: $B_{s}^{0}$ oscillation 
amplitude with statistical and systematic 
errors.
\label{fig:d0bsmix}}
\end{center}
\end{figure}

 Both CDF and D\O\ experiments have preliminary results on $B^{0}_{s}$ mixing measurements~\cite{bsmix}.
The D\O\ Collaboration used the semileptonic data sample corresponding to an
integrated luminosity of $460$~pb$^{-1}$.  About $13,300$ $B^{0}_{s}$ candidates have been reconstructed in 
the decay mode \mbox{$B^{0}_{s}\ra D_{s}^{-} \mu^{+} \nu, \ D_{s}^{-}\ra \phi\pi^{-}, \phi\ra K^{+}K^{-}$}. 
The opposite-side tagging technique was used to determine the initial state flavor of the $B^{0}_{s}$ meson.
A muon on the opposite side was always required. 
The muon charge information was combined with the muon 
transverse momenta (absolute and relative to the jet axis) and the opposite side
 secondary vertex charge in order to 
improve the tagging. The mistag rate has been determined from $\mu^{+} \overline{D^{0}}$ 
(mainly $B_{u}$ decays) and $\mu^{+} D^{*-}$ (mainly $B^{0}_{d}$ decays) samples to be equal to
$(26.6\pm 1.5)\%$ for $B_{u}$ and $(27.6\pm 2.1)\%$ for $B^{0}_{d}$ mesons. No $B^{0}_{s}$ oscillations 
were observed in the analysis
and an amplitude scan~\cite{bsscan} was used to determine a limit on the $B^{0}_{s}$ oscillation frequency.
Figure~\ref{fig:d0bsmix} shows the result of this scan which 
gave the $95\%$ C.L. limit of $5.0$~ps$^{-1}$ 
and sensitivity of $4.6$~ps$^{-1}$. 

\begin{figure}
\begin{center}
\epsfig{figure=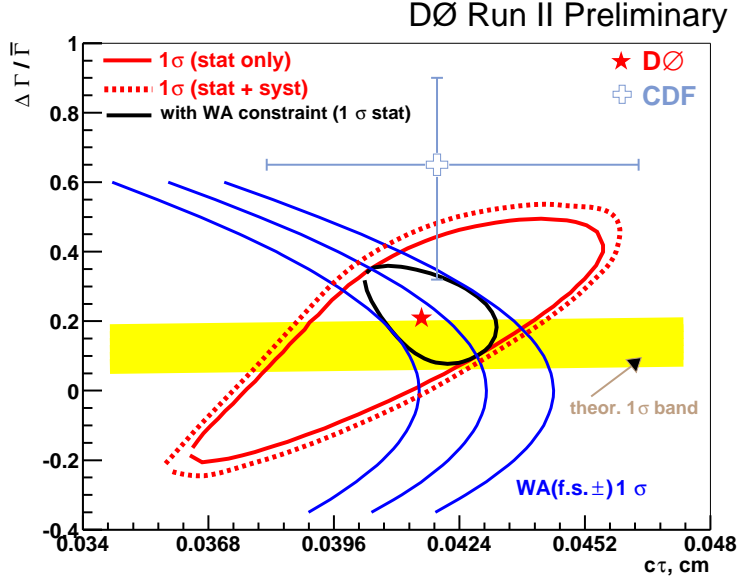,height=3in}
\caption{ 
The 1-$\sigma$ contour for the lifetime difference in the $B^{0}_{s} - \overline{B^{0}_{s}}$ system
from the fit to the D\O\ data, compared to a 1-$\sigma$ band 
for the World Average (WA) measurement based on semileptonic decays, 
$c\tau_{WA} = 430.0 \pm 15.0$~$\mu$m. A simultaneous fit to the data and WA 
gives a 1-$\sigma$ range $c\tau_{s} = 418_{-15}^{+14}$~$\mu$m and 
$\Delta\Gamma_{s}/\overline{\Gamma}_{s} = 0.23^{+0.16}_{-0.17}$ (stat.+syst.).
\label{fig:d0bsldiff}}
\end{center}
\end{figure}

 The CDF and D\O\ experiments used the $B^{0}_{s} \ra J/\psi \phi$ decays with no tagging requirement
 to determine the 
lifetime difference in the $B^{0}_{s} - \overline{B^{0}_{s}}$ system~\cite{cdfdgammas,d0dgammas}.
The D\O\ reconstructed
$483\pm 32$ $B^{0}_{s} \ra J/\psi(\ra \mu^{+}\mu^{-}) \phi(\ra K^{+}K^{-})$ candidates in 
$450$~pb$^{-1}$. The average lifetime of the $B^{0}_{s} - \overline{B^{0}_{s}}$ system was determined to be 
$\overline{\tau}_{s} = 1.39^{+0.15}_{-0.19}$~ps and the relative 
width difference 
$\Delta\Gamma_{s}/\overline{\Gamma}_{s}\equiv (\Gamma_{Ls} - \Gamma_{Hs})/\overline{\Gamma}_{s} = 0.21^{+0.33}_{-0.45}$.
The constraint from the world average (WA) of the $B^{0}_{s}$ lifetime measurements using semileptonic 
decays allows to decrease the relative width difference uncertainty, 
$\Delta\Gamma_{s}/\overline{\Gamma}_{s} = 0.23^{+0.16}_{-0.17}$. This result is in agreement with 
the CDF result~\cite{cdfdgammas} 
\mbox{$\Delta\Gamma_{s}/\overline{\Gamma}_{s} = 0.65^{+0.25}_{-0.33}\pm 0.01$} and 
theoretical predictions~\cite{dgammasth}
(see Fig.~\ref{fig:d0bsldiff}).

 The precisions of the $B^{0}_{s}$ mixing and lifetime difference measurements are statistically limited
for both detectors and one can expect significant improvements in the near future.

\section{Rare decays}

  Large $B$ and $D$ samples collected with the CDF and D\O\ detectors allow the search for and 
observation of rare beauty and charm decays.

 The CDF Collaboration 
used a $180$~pb$^{-1}$ data sample collected with the displaced track trigger for observation
of the $b\ra sss$ decays of $B^{0}_{s}$ mesons: 12 $B^{0}_{s}\ra \phi \phi$ candidate events have been 
observed with $1.95$ background events expected~\cite{cdfbsphiphi}. This gives the branching ratio
$BR(B^{0}_{s}\ra \phi \phi) = (1.4^{+0.6}_{-0.5}(stat.)\pm 0.6(syst.))\times 10^{-5}$ that is at the 
lower limit of the theoretical expectation in the Standard Model: 
$BR(B^{0}_{s}\ra \phi \phi) = (2.5 - 5.0)\times 10^{-5}$.

\begin{figure}
\begin{minipage}{0.47\textwidth}
\epsfig{figure=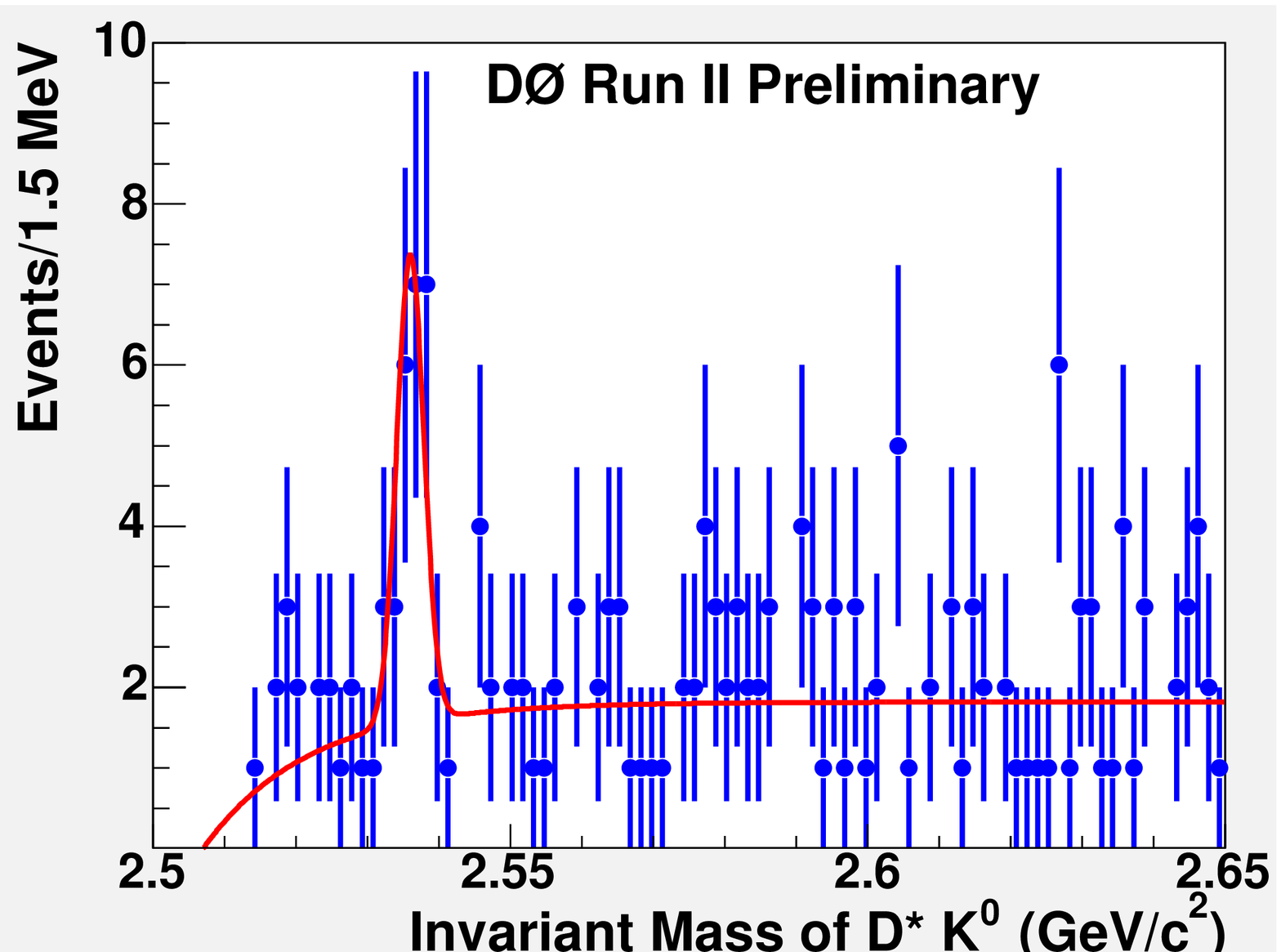,width=0.97\textwidth}
\caption{D\O\ Collaboration: Invariant mass of $D^{*}K^{0}_{s}$. 
Shown is the result of the fit of 
the $D^{*}K^{0}_{s}$ mass with a Gaussian function plus an exponential function 
with a threshold cut-off at $M(D^{*})+M(K^{0}_{s})$ to model the background. 
The total number of candidates in the peak is $18.5\pm 5.5$.
\label{fig:d0bsmuds1}}
\end{minipage}
\hfill
\begin{minipage}{0.47\textwidth}
\epsfig{figure=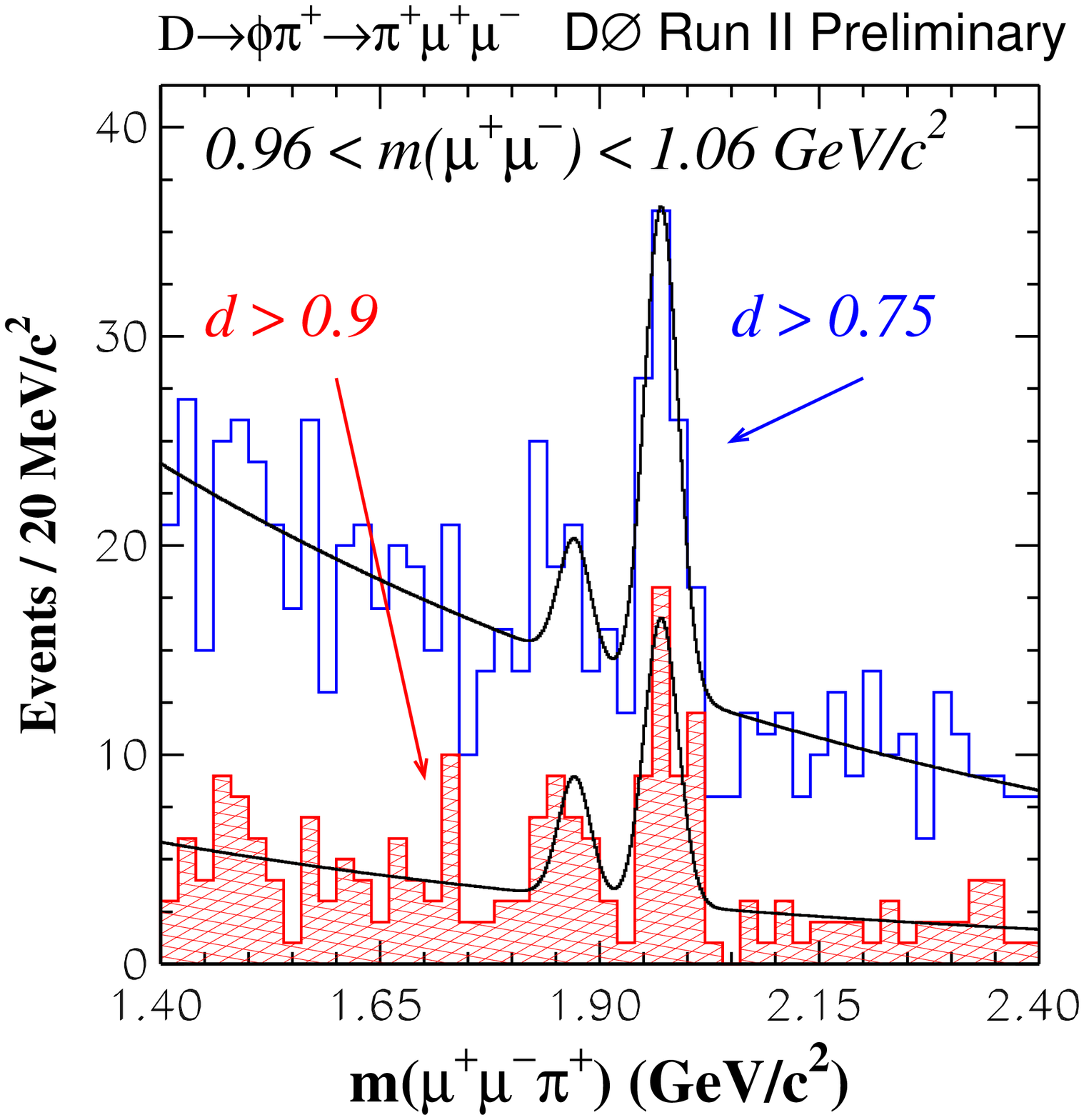,width=0.99\textwidth}
\caption{D\O\ Collaboration: The $m(\pi^{+}\mu^{+}\mu^{-})$ mass spectrum 
for events with the likelihood ratio requirement $d > 0.75$ (empty) and 
$d > 0.9$ (hatched). The results of binned likelihood fits to the 
distributions including contributions for $D^{+}_{s}$, $D^{+}$, and 
combinatoric background are overlaid on the histograms.
\label{fig:d0pimumu}}
\end{minipage}
\end{figure}

 The large coverage of the muon system in the D\O\ detector helped the observation of $18.5\pm 5.5$ 
$B^{0}_{s}\ra \mu D_{s1}(\ra D^{*\pm}K^{0}_{s}) X$ events in 
a $483$~pb$^{-1}$ data sample~\cite{d0bsds1mu} 
(see Fig.~\ref{fig:d0bsmuds1}). It is a first observation of this $B^{0}_{s}$ decay mode.

 Rare $B^{0}_{s}$ flavor-changing neutral current (FCNC) decays attract close attention 
from both theoretical and experimental sides. $B^{0}_{s}$
dimuon decays represent particular interest. The CDF and D\O\ collaborations already achieved the 
sensitivity that is enough to constrain some theoretical models predicting enhancement for 
such decays. Both CDF and D\O\ have results on the decay $B^{0}_{s}\ra \mu^{+} \mu^{-}\ $~\cite{bsmumu}. 
CDF constrained the branching ratio $BR(B^{0}_{s}\ra \mu^{+} \mu^{-}) < 7.5\times 10^{-7}$ at $95\%$ C.L. 
using a $171$~pb$^{-1}$ data sample and 
an update of the similar analysis
 from the D\O\ detector sets the upper limit $BR(B^{0}_{s}\ra \mu^{+} \mu^{-}) < 3.7\times 10^{-7}$ 
($95\%$ C.L.) with an integrated luminosity of $300$~pb$^{-1}$. 

Another FCNC decay, 
$B^{0}_{s}\ra \mu^{+} \mu^{-} \phi$ was searched for with the 
D\O\ detector~\cite{d0bsmumuphi}. The expected
sensitivity at $95\%$ C.L. has been determined: 
$\langle BR(B^{0}_{s}\ra \mu^{+} \mu^{-} \phi) \rangle = 1.2\times 10^{-5}$.
It can be compared with the upper limit $BR(B^{0}_{s}\ra \mu^{+} \mu^{-} \phi) < 6.7\times 10^{-5}$ 
set at $95\%$ C.L. with the CDF detector in Run I~\cite{cdfbsmumuphi}.

 The D\O\ Collaboration~\cite{d0dpimumu} searched for the FCNC charm decays $c\ra u \mu^{+} \mu^{-}$ 
using the data sample corresponding to $508$~pb$^{-1}$. 
The first step is an observation of the $D_{s}^{+} \ra \pi^{+} \phi(\ra \mu^{+}\mu^{-})$ decay 
(see Fig.~\ref{fig:d0pimumu}) which will be used as a normalization process in search for events
where the $\mu^{+}\mu^{-}$ system is not produced through a resonance.


\section*{References}
\begin{thebibliography}{99}

\bibitem{bfac} BABAR,  http://www.slac.stanford.edu/BFROOT/;\\
BELLE, http://belle.kek.jp/.

\bibitem{latQCD} M.~Wingate, Status of Lattice Flavor Physics (talk at Lattice2004), 
arXiv:hep-lat/0410008.\\
C.~Bernard's presentation at CKM2005, \\ http://ckm2005.ucsd.edu/agenda/wed1/bernard.pdf.

\bibitem{bd_freq} Heavy Flavor Averaging Group, Averages of b-hadron Properties as of Summer 2004,
arXiv:hep-ex/0412073.

\bibitem{bsmix} The CDF $B^{0}_{s}$ mixing result is described in the F.~Bedeschi's paper 
of these proceedings 
and at http://www-cdf.fnal.gov/physics/new/bottom/050310.bsmix-combined/. \\
The D\O\ $B^{0}_{s}$ mixing analysis is available at \\
http://www-d0.fnal.gov/Run2Physics/WWW/results/prelim/B/B20/.

\bibitem{bsscan} H.~G.~Moser and A.~Roussarie, \NIMA {\bf 384}, 491 (1997).

\bibitem{cdfdgammas} 
CDF Collaboration: D. Acosta {\it et al.}, {\em Phys. Rev. Lett.} {\bf 94} (2005) 101803.

\bibitem{d0dgammas}
D\O\ Collaboration: http://www-d0.fnal.gov/Run2Physics/WWW/results/prelim/B/B17/.

\bibitem{dgammasth} A.~Lenz, arXiv:hep-ph/0412007.

\bibitem{cdfbsphiphi} CDF Collaboration: D.~Acosta {\it et al.}, First Evidence 
for $B_s \to \phi \phi$ Decay and Measurements of 
Branching Ratio and $A_{CP}$ for $B^+ \to \phi K^+$, arXiv:hep-ex/0502044.

\bibitem{d0bsds1mu} D\O\ Collaboration: http://www-d0.fnal.gov/Run2Physics/WWW/results/prelim/B/B19/.

\bibitem{bsmumu} CDF collaboration:  D. Acosta {\it et al.}, \PRL {\bf 93} (2004) 032001. \\
D\O\ Collaboration: V.~M.~Abazov {\it et al.}, \PRL {\bf 94} (2005) 071802,\\
D\O\ update: http://www-d0.fnal.gov/Run2Physics/WWW/results/prelim/B/B21/.

\bibitem{d0bsmumuphi} D\O\ Collaboration: http://www-d0.fnal.gov/Run2Physics/WWW/results/prelim/B/B18/.

\bibitem{cdfbsmumuphi} CDF Collaboration: D.~Acosta {\it et al.}, \PRD 65 (2002) 111101.

\bibitem{d0dpimumu} D\O\ Collaboration: http://www-d0.fnal.gov/Run2Physics/WWW/results/prelim/B/B22/.

\end{thebibliography}
\end{document}